\documentclass{report} 
\usepackage{svcon2e}
\usepackage{graphicx}

\begin{document}
\chapter{On the statistics of the gravitational field}
\chapterauthors{A. Del Popolo}
\begin{abstract} 
In this paper we extend Chandrasekhar and von Neumann's analysis of the
statistics of the gravitational field to systems in which particles 
({\it e.g.} stars, galaxies) are not homogeneously distributed.  
We derive a
distribution function $ W({\bf F},d{\bf F}/dt)$ giving the joint
probability that a test
particle is subject to a force {\bf F} and an associated rate of change of 
{\bf F} given by d{\bf F}/dt. We calculate the first moment of d{\bf F}/dt 
to study the effects of inhomogenity on dynamical friction. 
\end{abstract} 
\section{Introduction}
The study of the statistics of the fluctuating gravitional force in infinite 
homogeneous systems was pionered by Chandrasekhar \& von
Neumann in two classical papers (Chandrasekhar \& von Neumann
1942, 1943 hereafter CN43). The analysis of
the fluctuating gravitional field, developed by the quoted authors, was
formulated by means of a statistical treatment in the case of uniform systems, 
and with no correlations.
 Two distributions are fundamental for the description of the fluctuating 
gravitional field: 
\begin{itemize}
\item 1) $ W({\bf F})$ which gives the probability that a test star
is subject to a force ${\bf F}$ in the range ${\bf F}$, ${\bf F}$+
d${\bf F}$;
\item 2) $ W({\bf F},{\bf f})$ which gives the
joint probability that the star experiences a force {\bf F} and a rate 
of change {\bf f}, where $ {\bf f} = d{\bf F}/dt $.
\end{itemize}
From a pure theorethical ground we expect that inhomogeneity affects all the aspects of the fluctuating gravitational field (Antonuccio \& Colafrancesco 1994; Del Popolo 1996a, b; Del Popolo \& Gambera 1998).
\section{$ W({\bf F},{\bf f})$ and $ \overline f$ in inhomogeneous systems} 
Assuming that the system density is described by a powe law of index $p$, the expression of $ W({\bf F},{\bf f})$ is given following Markoff's
method by (CN43):
\begin{eqnarray}
W({\bf F},{\bf f}) & = & \frac{1}{64 \pi^{6}} \int_{0}^{\infty}
\int_{0}^{\infty}  A({\bf k},{\bf \Sigma}) \cdot  \nonumber \\
                   &   &\left\{ \exp{[-i({\bf k}{\bf \Phi} +
{\bf \Sigma}{\bf \Psi})]} \right\} d{\bf k} d{\bf \Sigma}
\label{eq:set}
\end{eqnarray}
A lenghty calculation leads us (see Del Popolo \& Gambera 1998 for a derivation and the
meaning of simbols) to find the function
$A({\bf k}, {\bf \Sigma})$:
\begin{eqnarray}
A({\bf k}, {\bf \Sigma}) & = & e^{  -\tilde{a} k^{\frac{3-p}{2}}} \{1- \; i g p({\bf k}, {\bf \Sigma})   \nonumber \\
                            &   & + \; \tilde{b} k^{\frac{- (3 
+ p)}{2}} ] \cdot  [ Q({\bf \Sigma}) + k R({\bf \Sigma}) ] \} 
\label{eq:trense}
\end{eqnarray} 
This last equation introduced into Eq. (\ref{eq:set}) solves the problem
of finding the distribution $W(\bf F,\bf f)$ and makes it possible to find
the moments of $\bf f$ that give information regarding the dynamical
friction.
Using this last expression for $A({\bf k}, {\bf \Sigma})$ and performing a calculation similar to that by CN43 the first moment of  $ {\bf f}$ is given by:
\begin{eqnarray}
\overline {\bf f} & = & -\left( \frac{1}{2} \right)^{\frac{3}{3 - p}} \cdot A(p) \cdot B(p)^{\frac{p}{3 - p}} \nonumber \\
                  &   &\cdot \frac {\alpha^{\frac{3}{3 - p}} G M L(\beta)}{ \pi H(\beta) \beta^{\frac{2 - p}{2}}} \cdot \left[ {\bf v} - \frac{3 {\bf F} \cdot {\bf v}}{ |  {
\bf F} |^2} \cdot {\bf F} \right]
\label{eq:quaruno}
\end{eqnarray}
where
\begin{eqnarray}
L ( \beta) & = & 6 \int_{0}^{\infty} \left[
e^{(x/\beta)^{\frac{(3 - p)}{2}}} \right]
\left[ \frac{ \sin{x}}{x^{(2 - p)/2}} -
\frac{\cos{x}}{x^{p/2}} \right] dx \nonumber \\
           &   & - \; 2 \int_{0}^{\infty}
           \left[ e^{(x/\beta)^{\frac{(3 - p)}{2}}}
           \right] \cdot \frac{\sin{x}}{x^{(p - 2)/2}} dx
\label{eq:quardu}
\end{eqnarray} 
As shown by Eq. (\ref{eq:quaruno}), in a inhomogeneous system, differently from homogeneous ones, $ {\bf f}$ 
is a function of the inhomogeneity parameter
$p$. 

At this point we may show how dynamical friction changes due to 
inhomogeneity. From Eq. (\ref{eq:quaruno}) we see that
$ \frac{d{\bf F}}{dt}$ differs from that obtained in homogeneus system 
only for the presence of a dependence on the inhomogeneity parameter 
$ p$. If we divide Eq. (\ref{eq:quaruno}) for the correspondent of CN43 we 
obtain:
\begin{eqnarray}
\frac{ \left( \overline{\frac{d{\bf F}}{dt}} \right)_{Inh.}}
{ \left( \overline{\frac{d{\bf F}}{dt}} \right)_{Hom.}} 
& = & \; - \left( \frac{1}{2} \right)^{\frac{6 - p}{3 - p}} \cdot  \frac{3 \alpha^{\frac{3}{3 - p}} L(\beta) B(p)^{\frac{p}{3 - p}} \cdot A(p) }{n \cdot \pi^{2} H(\beta) B(\beta) \beta^{\frac{2 - p}{2}}}
\label{eq:cinqsei}
\end{eqnarray}
This last  equation is an increasing function of $p$. This means
that for increasing values of $ p$ 
the star suffers an even greater amount of acceleration in the
direction $ - {\bf v}$ (when
$ {\bf v} \cdot {\bf F} \le 0$) than in the direction $ +{\bf v}$ (when 
$ {\bf v} \cdot {\bf F} \ge 0$), with respect to the homogeneous case.
This is due to the fact that the difference
between the amplitude of the decelerating impulses and the accelerating ones 
is, as in homogeneous systems, statistically negative, but now larger,
being the scale factor greater.
This finally means that, for a given value of $n$, 
the dynamical friction increases with increasing inhomogeneity in
the space distribution of stars. 
%
In addition, by 
increasing $n$ the dynamical friction increases, just like in the 
homogeneous systems, but the increase is larger than the 
linear increase observed in homegeneous ones.

\end{document}